\documentclass[twocolumn,twoside,slac_two]{revtex4-1}
\usepackage{graphicx}
\usepackage{fancyhdr}
\usepackage{hyperref}
\hypersetup{
    colorlinks=true,
    urlcolor=magenta,
}
\begin{document}
\title{Indications for a cascade component in $\gamma$-ray blazar spectra}
%
\author{S. Baklagin, T. Dzhatdoev, E. Khalikov}
\affiliation{Federal State Budget Educational Institution of Higher Education M.V. Lomonosov Moscow State University, Skobeltsyn Institute of Nuclear Physics (SINP MSU), 1(2), Leninskie gory, GSP-1, Moscow 119991, Russian Federation; e-mail: timur1606@gmail.com, nanti93@mail.ru}
\author{A. Kircheva}
\affiliation{Federal State Budget Educational Institution of Higher Education M.V. Lomonosov Moscow State University, Department of Physics, 1(2), Leninskie gory, GSP-1, Moscow 119991, Russian Federation and Federal State Budget Educational Institution of Higher Education M.V. Lomonosov Moscow State University, Skobeltsyn Institute of Nuclear Physics (SINP MSU), 1(2), Leninskie gory, GSP-1, Moscow 119991, Russian Federation}
\author{A. Lyukshin}
\affiliation{Federal State Budget Educational Institution of Higher Education M.V. Lomonosov Moscow State University, Department of Physics, 1(2), Leninskie gory, GSP-1, Moscow 119991, Russian Federation}
\begin{abstract}
We present a very brief overview of some recent $\gamma$-ray observations of selected blazars to reveal an indication for a considerable or even dominant contribution of secondary $\gamma$-rays from electromagnetic cascades to the observable spectra in the 1--500 $GeV$ energy range.
\end{abstract}
\maketitle
\thispagestyle{fancy}
\section{INTRODUCTION}

The Universe is filled by light. There are two principal sources of such photons --- the cosmic microwave background (CMB) and the extragalactic background light (EBL). While CMB properties are quite well measured and theoretically understood for more than 20 years (e.g. \cite{r1}), many EBL models with widely different parameters were developed \cite{r2}--\cite{r10}; for instance, the total EBL intensity for these models differs by more than a factor of two \cite{r11}. Comparatively recently, however, some of these EBL models started to converge, at least in the 0.5--10 $nm$ wavelength region (e.g., \cite{r4} and \cite{r9}). 

High-energy $\gamma$-rays with primary energy $E_{0}>0.25(1 [eV]/\epsilon) [TeV]$ get absorbed \cite{r12}--\cite{r13} ($\gamma\gamma\rightarrow e^{+}e^{-}$) on EBL and CMB photons with energy $\epsilon$, reaching maximum cross section at $E_{0}\approx (1 [eV]/\epsilon) [TeV]$ \cite{r9}. Therefore, the improved reliability of EBL models allows for more detailed studies of this fundamental quantum electrodynamics process in the $\gamma$-ray energy range of about 500 $GeV$-- 10 $TeV$, and of extragalactic $\gamma$-ray propagation effects in general.

Using the CMB for such studies is, in principle, also possible \cite{r14}, but at present is not feasible due to several factors. Indeed, the typical mean free path $L$ of a $\sim10^{15}$ $eV$= 1 $PeV$ $\gamma$-ray on the CMB is $\sim$10 $kpc$; for such high energies there is still no discovered sources, while using lower energies is not convenient either as in the $E_{0}<$100 $TeV$ energy region the $L(E_{0})$ dependence is very strong: $L$ falls for about an order of magnitude for every 10 \% of $E$ decrease. Therefore, extremely well knowledge of experimental systematics on primary $\gamma$-ray energy would be crucial while using CMB photons at $E_{0}<$100 $TeV$ as a probe of the $\gamma\gamma\rightarrow e^{+}e^{-}$ process.

In the present conference contribution we emphasize on the potential importance of the development of electromagnetic cascades on the EBL/CMB to the interpretation of observations in the high energy (HE, $E$=100 $MeV$--100 $GeV$) and the very high energy (VHE, $E$=100 $GeV$--100 $TeV$) ranges. This short paper does not pretend to claim great originality; most of its basic ideas were already discussed and most of its main results were already presented in \cite{r15}--\cite{r19}.

\section{EXTRAGALACTIC $\gamma$-ray PROPAGATION: ANOMALIES AND MODELS}

\subsection{Absorption-Only Model}
\begin{figure*}[t]
\centering
\includegraphics[width=110mm]{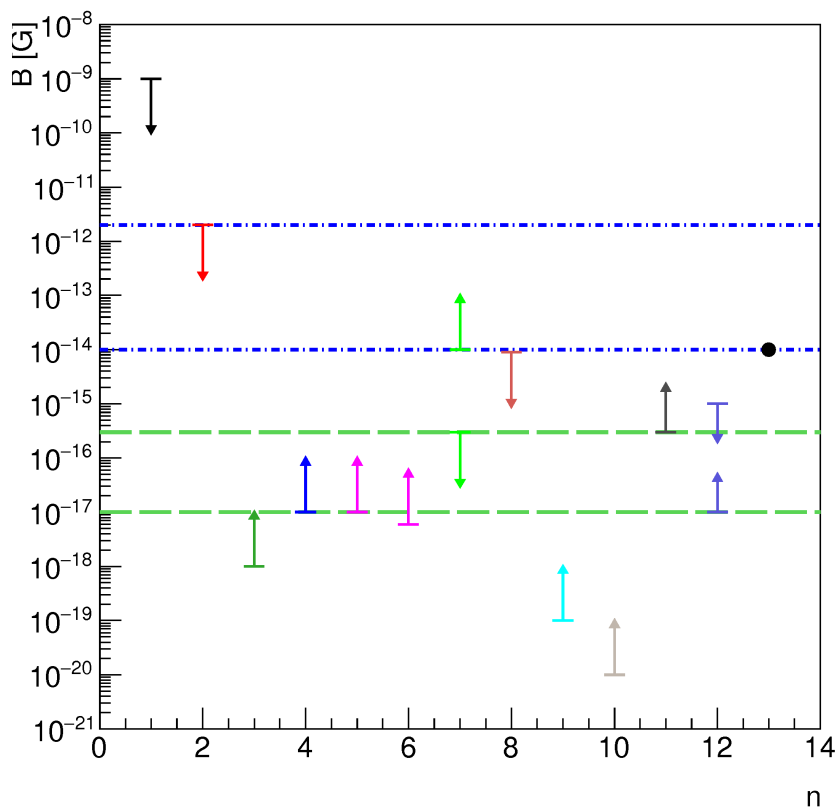}
\caption{Some constraints on the EGMF strength in voids (a figure from \cite{r19}).} \label{f1}
\end{figure*}
Soon after the discovery of the first $TeV$ extragalactic $\gamma$-ray source \cite{r20}, the first $\gamma$-astronomical constraints on the EBL density were obtained \cite{r21}--\cite{r22}, accounting for only the pair-production process as the main cause of transformation of the primary $\gamma$-ray spectrum (for very distant sources, it is necessary to include adiabatic losses). This model we call ``the absorption-only model''. Since 1993, almost every year several papers appear, assuming the absorption-only model, sometimes at a rate of a dozen a year or more.
\subsection{Anomalies}
In what follows, any significant deviations from the absorption-only model are called ``anomalies'', even if they do not fall beyond the conventional physics. Several such anomalies were reported \cite{r23}--\cite{r27}, with a statistical significance not overwhelming in every single case, but together they indicate that the absorption-only model is incomplete and must be modified in some way (see \cite{r15} for more details).

\subsection{Extragalactic Magnetic Field}
Secondary electrons and positrons (hereafter simply ``electrons'') produced in the $\gamma\gamma\rightarrow e^{+}e^{-}$ process radiate cascade $\gamma$-rays by means of the Inverse Compton (IC) process; these $\gamma$-rays may contribute to the observable spectrum of an extragalactic source. Therefore, besides the properties of background photon fields (EBL/CMB), another very important factor is the strength and structure of the extragalactic magnetic field (EGMF). Some selected constraints on the EGMF strength $B$ in voids of the large scale structure (LSS), most of them assuming correlation length of the field $L_{c}$= 1 $Mpc$, are shown in Figure~\ref{f1}. There are two regions of $B$ values that satisfy most of these constraints: $10^{-14}$ $G<B<$2$\cdot 10^{-12}$ $G$ and $10^{-17}$ $G<B<$3$\cdot 10^{-16}$ $G$. The first case corresponds to the absorption-only model regime: for such a strong EGMF secondary electrons are, as a rule, strongly deflected and delayed so that cascade photons do not contribute to the point-like image of the source. On the other hand, for the case of the second option such a contribution is still possible, at least in the VHE energy range. The lower bound on the $B$ value, $10^{-17}$ $G$, is highly uncertain \cite{r28}-\cite{r29}. Very recently a paper \cite{r30} appeared, disfavouring a narrow range of values around $B$= $10^{-14}$ $G$ for $L_{c}$= 1 $Mpc$.

\subsection{Electromagnetic Cascade Model}
\begin{figure*}[t]
\centering
\includegraphics[width=110mm]{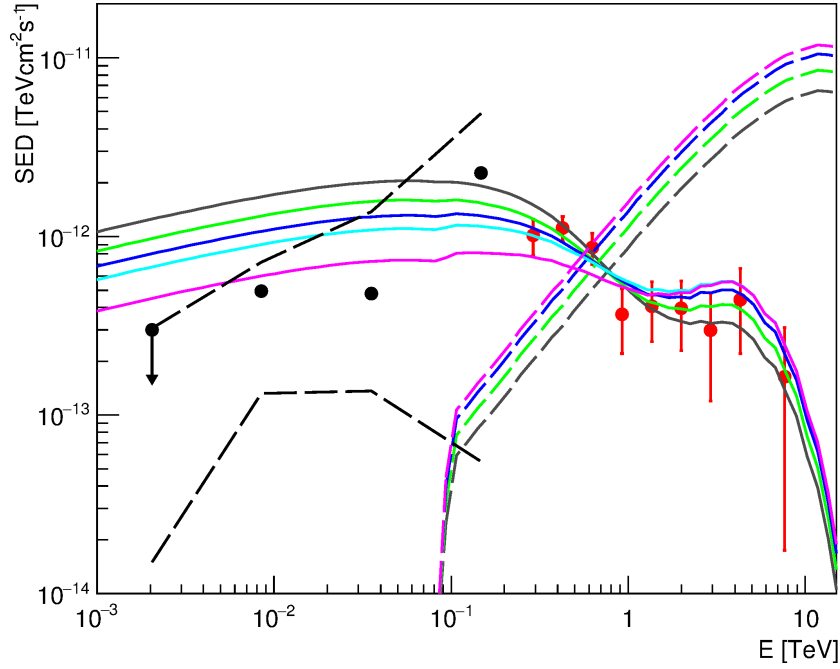}
\caption{Several fits (solid curves) to the spectrum of blazar 1ES 0229+200 \cite{r31}--\cite{r32} obtained in the framework of the intergalactic electromagnetic cascade model. Black curve denotes $V$= 1, green curve --- $V$= 0.6, blue --- $V$= 0.4, cyan --- $V$= 0.3, magenta --- $V$= 0.2. Primary (intrinsic) spectra for each case are also shown by the same colors (dashed curves in the right part of the graph). Full uncertainty for the four lowest-energy bins is shown by black dashed lines.} \label{f2}
\end{figure*}
\begin{figure*}[t]
\centering
\includegraphics[width=110mm]{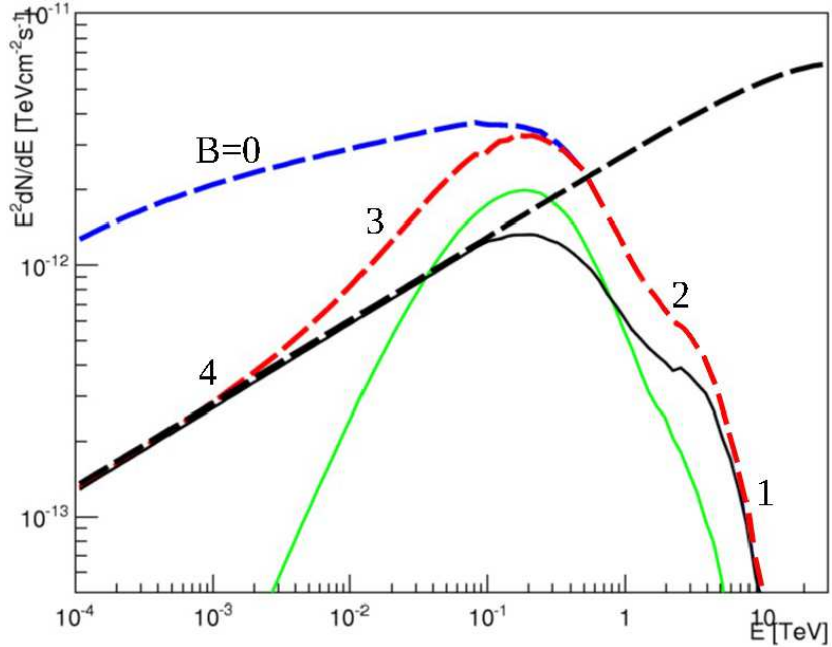}
\caption{The main spectral signatures of the electromagnetic cascade model (a figure from \cite{r35}).} \label{f3}
\end{figure*}

Let us assume for a moment the following simple two-phase model of the EGMF. LSS voids with $B=0$ fill a fraction of the total volume $0<V<1$, while the rest is occupied by the comparatively strong ($B>10^{-14}$ $G$) EGMF with $L_{c}\sim1$ $Mpc$. Figure~\ref{f2} shows several fits to the observed spectral energy distribution (SED) of blazar 1ES 0229+200 (\cite{r31} for the four lowest-energy bins shown in black and \cite{r32} for other bins) assuming this model with various values of $V$. Cascade component dominates at low energies ($E<$300--500 $GeV$). Details of simulations correspond to the case of Figure 15 of \cite{r15}. Reasonable fits in the energy range 30 $GeV$-- 10 $TeV$ may be obtained for the case of $V$ between 0.4 and 0.6. However, below 30 $GeV$ the model intensity exceeds the observed one. This may be explained either by the deflection and/or delay of cascade electrons while travelling the EGMF \cite{r33}--\cite{r34} or by additional (with respect to IC) electron energy losses \cite{r29}.

Figure~\ref{f3} depicts the main spectral signatures of the electromagnetic (EM) cascade model: 1) high-energy cutoff, similar to the absorption-only model, 2) an ankle at the intersection of the primary and cascade components (in fact, this signature is also visible in Figure~\ref{f2}), 3) a low-energy cutoff for the case of a non-zero EGMF value (``magnetic cutoff''), and 4) a ``second ankle'' at the low-energy region of the spectrum.

\subsection{Explanation of anomalies in the EM Cascade Model}

We argue that all above-mentioned anomalies may be interpreted within the framework of the EM cascade model, namely: \\
1. The apparent excess of observed $\gamma$-rays in the highest-energy bins claimed in \cite{r23}--\cite{r24} was discussed by \cite{r15}. A prominent ankle (signature 2 in Figure~\ref{f3}) may account for this effect. \\
2. The unusual spectral hardening towards lower energies \cite{r25} is explained by the existence of a ``magnetic cutoff'' (signature 3 in Figure~\ref{f3}), as the authors of \cite{r25} themselves note (however, see \cite{r36} and \cite{r25} itself for other possible explanations). \\
3. The same spectral feature (a prominent magnetic cutoff) may account for the result of \cite{r26}, as discussed in \cite{r35}. \\
4. Finally, ``halos'' observed around some blazars also may be explained in the context of the EM cascade model \cite{r27}. \\
Thus, the EM cascade model appears to be the simplest extragalactic $\gamma$-ray propagation scenario that coherently explains all these anomalies. Other, more exotic models that could account for a part of these anomalies do exist \cite{r37}--\cite{r38}; the main difficulties of these scenarios were discussed by us in \cite{r15},\cite{r18}.

\section{CONCLUSIONS}

Recent observations of some blazars in the 1 $GeV$ -- 10 $TeV$ energy region seem to indicate that the secondary component of cascade $\gamma$-rays may constitute a considerable contribution to the observable flux in the $E=$1--500 $GeV$ energy range.
\bigskip 
\begin{acknowledgments}
This work was supported by the RFBR Grant 16-32-00823. T.D. acknowledges the support of the Students and Researchers Exchange Program in Sciences (STEPS), the Re-Inventing Japan Project, JSPS, and the hospitality of the University of Tokyo ICRR.
\end{acknowledgments}

\bigskip 

\end{document}